\documentstyle[aps,prl,preprint,floats,epsf]{revtex}
\def\piz{\pi^0}
\def\B{{\cal B}}
\def\eff{\varepsilon}
\def\thrhgg{$\pi^-\pi^+\pi^-\eta_{\gamma\gamma}$}
\def\thrh3p{$\pi^-\pi^+\pi^-\eta_{3\piz}$}
\def\h2pgg{$\pi^-2\piz\eta_{\gamma\gamma}$}

\input psfig

\input epsf

\newcommand{\pssize}[2]{
\vspace{-80pt}
\vspace{+15pt}
  \setlength{\epsfysize}{#2truein}
\centerline{\epsfbox{#1}}\vspace{-15pt}
}

\newcommand{\Caption}[2]{

{{FIG. #1.}  #2}
}

\long\def\single#1#2#3#4#5{
\begin{figure}[#2]   
  \hskip -1cm\vskip +2.0cm\pssize{#3}{#4} 
  \hskip -2cm\vskip -0.0cm \Caption{#1}{#5}
  \vskip -0.0cm 
\end{figure}
}

\long\def\Ssingle#1#2#3#4#5{
\begin{figure}[#2]   
\vskip 0.5cm\hspace{ -3cm}
\pssize{#3}{#4} 
  \hskip -0.5cm\vskip  0.3cm \Caption{#1}{#5}
  \vskip -0.0cm 
\end{figure}
}

\long\def\Sfingle#1#2#3#4#5{
\begin{figure}[#2]   
  \hskip -1cm\vskip +2.0cm\pssize{#3}{#4} 
  \hskip -2cm\vskip    0.0cm \Caption{#1}{#5}
  \vskip -0.0cm 
\end{figure}
}

\long\def\Single#1#2#3#4#5{
\begin{figure}[#2]   
  \hskip -1cm\vskip +1.7cm\pssize{#3}{#4} 
  \hskip -2cm\vskip -0.0cm \Caption{#1}{#5}
\end{figure}
}

\begin{document}
\preprint{\tighten\vbox{\hbox{\hfil CLNS 97/1489 \hfill}
                        \hbox{\hfil CLEO  97-13 \hfill}
}}

\title{\quad\\ \Large
First observation of $\tau\to 3\pi\eta\nu_{\tau}$ and $\tau\to
f_1 \pi\nu_\tau$ decays.}

\author{CLEO Collaboration}
\author{(\today)}

\maketitle
\tighten

\begin{abstract}
We have observed new channels for $\tau$ decays with an $\eta$ in the final 
state. We study 3-prong tau decays, using the
$\eta\rightarrow\gamma\gamma$ and $\eta\rightarrow 3\piz$ decay modes
and 1-prong decays with two
$\piz$'s using the $\eta\rightarrow\gamma\gamma$ channel.
The measured branching fractions are 
$\B(\tau^{-}\to \pi^{-}\pi^{-}\pi^{+}\eta\nu_{\tau})=(3.4^{+0.6}_{-0.5}\pm0.6)\times10^{-4}$ 
and $\B(\tau^{-}\to \pi^{-}2\piz\eta\nu_{\tau})=(1.4\pm0.6\pm0.3)\times10^{-4}$. 
We observe clear evidence for $f_1\to\eta\pi\pi$ substructure and measure
$\B(\tau^{-}\to f_1\pi^{-}\nu_{\tau})=(5.8^{+1.4}_{-1.3}\pm1.8)\times10^{-4}$. 
We have also searched for $\eta'(958)$ production and obtain 90\% CL upper 
limits
$\B(\tau^{-}\to \pi^{-}\eta'\nu_\tau)<7.4\times10^{-5}$ and 
$\B(\tau^{-}\to \pi^{-}\piz\eta'\nu_\tau)<8.0\times10^{-5}$.  
\end{abstract}

\newpage

{
\renewcommand{\thefootnote}{\fnsymbol{footnote}}
\begin{center}
T.~Bergfeld,$^{1}$ B.~I.~Eisenstein,$^{1}$ J.~Ernst,$^{1}$
G.~E.~Gladding,$^{1}$ G.~D.~Gollin,$^{1}$ R.~M.~Hans,$^{1}$
E.~Johnson,$^{1}$ I.~Karliner,$^{1}$ M.~A.~Marsh,$^{1}$
M.~Palmer,$^{1}$
% ...
\end{center}
\newpage 
\small
\begin{center}
$^{1}${University of Illinois, Champaign-Urbana, Illinois 61801}\\
% ...
\end{center}

\setcounter{footnote}{0}
}
\newpage

Tau decays with an $\eta$ meson in the final state provide important
information about 
various hadronic symmetries
and allow for a study of the resonant structure of the weak hadronic
current.
These decays are rare and their detection became possible only recently    
with the high statistics
CLEO experiment at Cornell Electron Storage Ring(CESR).
Two such decays with small branching fractions already have been observed:
$\B(\tau^{-}\to\pi^{-}\piz\eta\nu_{\tau})=(0.17\pm0.02\pm0.02)\%$
\cite{etarho} and $\B(\tau^{-}\to K^{-}\eta\nu_{\tau})=(0.026\pm0.005\pm0.005)\%$
\cite{CLEO96-5}\cite{charge}.
Both channels also  have been seen by the ALEPH group \cite{alep2}.
 All other tau decays involving $\eta$ mesons were expected 
to be severely suppressed. 
The decay $\tau \to 3 \pi\eta\nu_{\tau}$ 
can proceed through the axial-vector current and its branching
fraction was predicted to be $1.2\times 10^{-6}$ \cite{pich}.

In this Letter, we present the first observation of the tau decay 
$\tau\to 3\pi\eta\nu_{\tau}$ using three final states:
\thrhgg\, where $\eta$ is reconstructed from the
$\eta\to\gamma\gamma$ decay; 
\thrh3p\, where $\eta$ is reconstructed from
its $\eta\to 3\piz$ decay; and  \h2pgg\ where $\eta$ is
reconstructed from the $\eta\to\gamma\gamma$ decay and the remaining photons 
form 2 $\piz$'s.  In addition, for the first time, we observe
$\tau^-\to f_1\pi^-\nu_{\tau}$ using the $f_1\to\eta\pi^+\pi^-$ decay mode. 
We also search for decays with $\eta'(958)$ 
using the $\eta'\to\eta\pi^+\pi^-$ decay mode with $\eta\to\gamma\gamma$.

We use data  obtained by the CLEO II detector \cite{cleo}
at the CESR operating at a center of mass energy 
corresponding to  the peak of the $\Upsilon(4S)$
resonance ($E_{cm}=$10.6 GeV) and 60 MeV below this energy. 
The data correspond to an integrated luminosity of 4.68 $fb^{-1}$ 
and contain about
4.27 million $\tau^{+}\tau^{-}$ pairs. CLEO II is a general
purpose solenoidal spectrometer.
In addition to good quality tracking,
its special feature is a 7800 crystal CsI(Tl) electromagnetic calorimeter 
that provides photon detection with high efficiency and good energy and angular
resolution, which is essential for $\eta$ and $\pi^0$ reconstruction.

We select events using the 1 vs 3 and 1 vs 1 charged track topologies and
tag one of the tau decays with a single charged track in the drift chamber 
which is required to be identified as an electron, muon or hadron. 
The electron candidate is required to have momentum,
$p$, greater than 0.5 GeV and energy deposition in the calorimeter, $E$, such 
that $0.9<E/p<1.1$. If specific ionization ($dE/dx$) information is available,
we veto the event if it is more than two standard deviations below the expected 
value. Muon candidates must penetrate at least 3 absorption lengths of material
for track momenta less than 2.0 GeV, and more than 5 absorption 
lengths for momenta above 2.0 GeV. 

A hadron tag is a track not identified as
an electron or muon and with momentum
pointing to the barrel part of the calorimeter, $|\cos\theta|<0.81$,
where $\theta$ is the polar angle defined with respect to the beam direction.
The invariant mass, including all photon candidates in the tag 
hemisphere, is required to be less than 1.2 GeV. In addition to single pions,
this tag recovers unidentified electrons and muons and a large fraction
of $\tau^-\to\rho^-\nu_\tau$ decays.
 
The second tau -- representing the signal candidate -- is reconstructed
from its decays into \thrhgg\ , \thrh3p\ and \h2pgg\ final states.
We assume that all charged tracks are pions since there is very 
little phase space for decays in which one of the tracks is a kaon.
The dE/dx information is consistent with this assumption.

Photons are identified by isolated energy clusters in the calorimeter,
separated from energy deposits left by
charged tracks and with photon-like lateral profiles of energy deposition. 
Photon candidates used for $\piz$ and $\eta$ reconstruction
are required to be in the barrel part of the calorimeter  and 
to satisfy $|S^X_{\gamma\gamma}|<10$, where
$S^X_{\gamma\gamma}\equiv({m_{\gamma\gamma}-m_X})/\sigma_{\gamma\gamma}$
($X=\piz$ or $\eta$) and $\sigma_{\gamma\gamma}$ is the $\piz$ or $\eta$ 
mass resolution($\sim$12 MeV). Only photon pair combinations with 
$-3.0<S^{\piz}_{\gamma\gamma}<2.0$ are considered as signal candidates; 
those with larger values of $|S^{\piz}_{\gamma\gamma}|$ are used for sidebands. 

For lepton (hadron) tags, the lower energy photon used for $\eta$
reconstruction must have energy greater than 200 (250) MeV and the higher
energy photon must have energy greater than 400 (700) MeV.  Photons used 
to form $\piz$'s are required to have energies greater than 30 MeV.
In events for which more than one combination of photons passes all cuts, we 
choose the combination with the smallest $\chi^2$ for that signal hypothesis.  

We  suppress $e^{+}e^{-}\to e^{+}e^{-}(\gamma)$
and $e^{+}e^{-}\to \mu^{+}\mu^{-}(\gamma)$ events
by vetoing events with tracks which have energy greater than 85$\%$ of the 
beam energy.
To remove background due to 2-photon processes,
we require the missing momentum vector of the event to 
be in the angular region $|\cos\theta|<0.9$.
We suppress contributions from tau decays with a $K_S$ in the final state
by requiring that for all tracks, the impact parameter with respect to the 
interaction point must be less than 5 mm.
Background from low multiplicity $q\bar{q}$ events and incompletely 
reconstructed tau events is minimized by rejecting events with additional
isolated photons with an energy greater than 120 MeV.
To further reduce $q\bar{q}$ background, we require the total invariant 
mass of the hadrons in the signal hemisphere to be less than the tau mass.
 
For the \thrhgg\ sample we reduce $q\bar{q}$  and 2-photon
backgrounds by requiring the event to have missing mass 
satisfying $0.1<M_{miss}/E_{cm}<0.5$ and 
total transverse momentum greater than 0.3 GeV \cite{Mmiss}.
Decays with $K_S$'s are additionally suppressed by requiring both $\pi^+\pi^-$
combinations to have a mass at least 15 MeV from the $K_S$ mass. 
To suppress background from events with gamma conversions,
we veto events with electron candidates in the signal hemisphere.

We simulate tau signal and background events using the 
{\tt KORALB} generator and {\tt TAUOLA} decay packages \cite{koralb}
(with some modifications discussed below) and measured tau branching
 fractions \cite{PDG}; {\tt GEANT} \cite{geant} is used for detector simulation. 

We find background associated with various QED processes to be negligible
by using Monte Carlo (MC) simulations and independent data samples.
To estimate $q\bar{q}$ background, we use independent data samples
requiring the invariant mass of the tag hemisphere to be greater than $1.8$ GeV.
We select predominantly hadronic events satisfying the same topological and 
kinematic requirements on the signal hemisphere as described above,
except for the tau mass cut.
The  normalization for this hadronic sample is obtained from a fit
to the data in the region with signal invariant mass above 1.8 GeV. 
  
\single{1}{h}{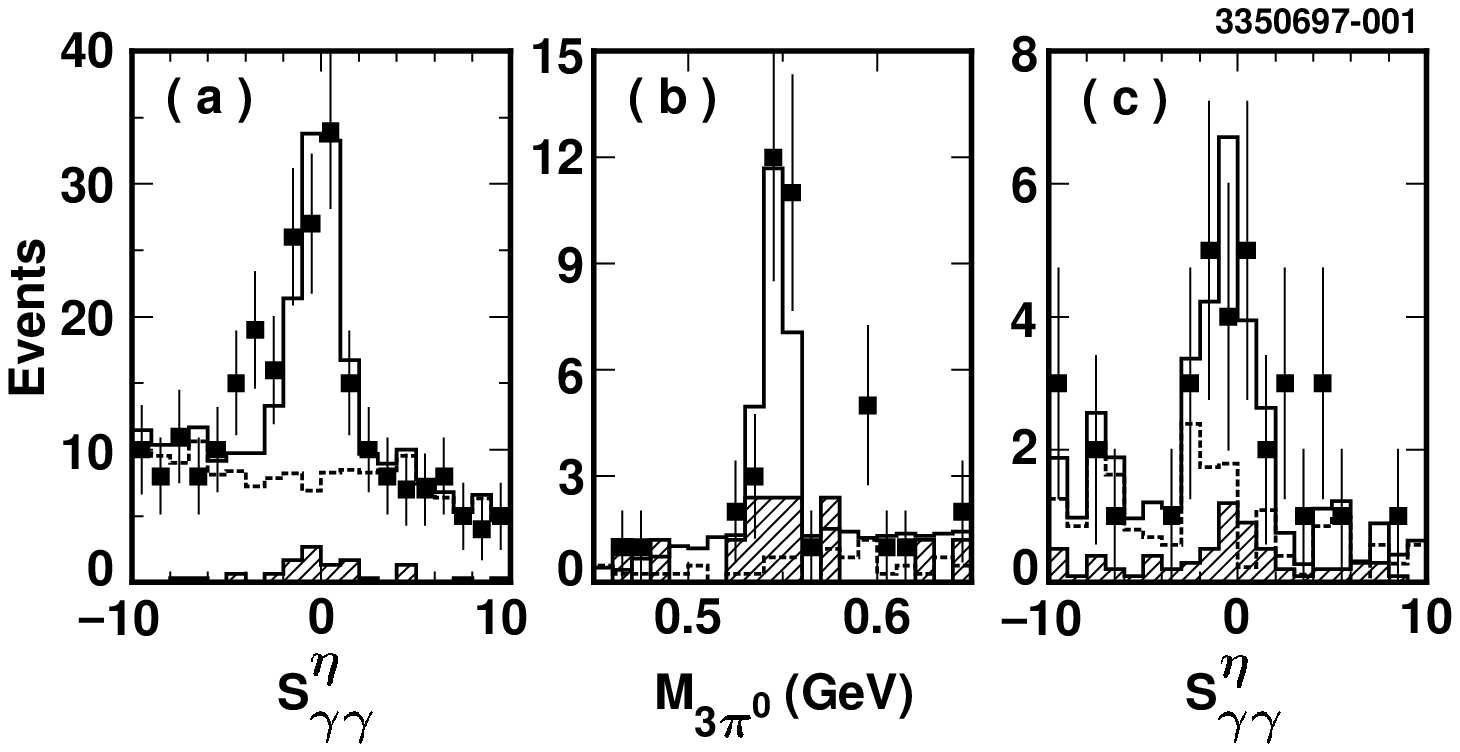}{2.5}{
Distribution of $\eta$ mass  for the (a) {\thrhgg}, (b) {\thrh3p} and (c) \h2pgg\  samples.
 The solid line is a fit to the data (squares).  The tau and $q\bar{q}$  backgrounds are 
indicated by the dashed line and hatched area, respectively.  Plots (a) and
(c) are binned in units of $S_{\gamma\gamma}$ while (b) has a
10 MeV bin size.  In (b), the energies and angles of each photon of the
$\piz$ candidates have been constrained to $\piz$ mass.}

The distributions illustrating $\eta$ signal in all three analyses are shown 
in  Fig. 1. Tau background contributes
almost entirely through random $\gamma\gamma$ combinations while
most of the coherent $\eta$ background comes from $q\bar{q}$ events. 
Distributions of hadronic masses for events from the $\eta$ signal regions
are shown in Fig. 2, where the events with masses above the tau mass are 
plotted as well. Signal regions for $\eta\to\gamma\gamma$ and $\eta\to3\piz$ channels
are  $-3.0<S^{\eta}_{\gamma\gamma}<2.0$ 
and $|M(3\piz)-M_{\eta}|<20$ MeV respectively.

\nopagebreak

\single{2}{h}{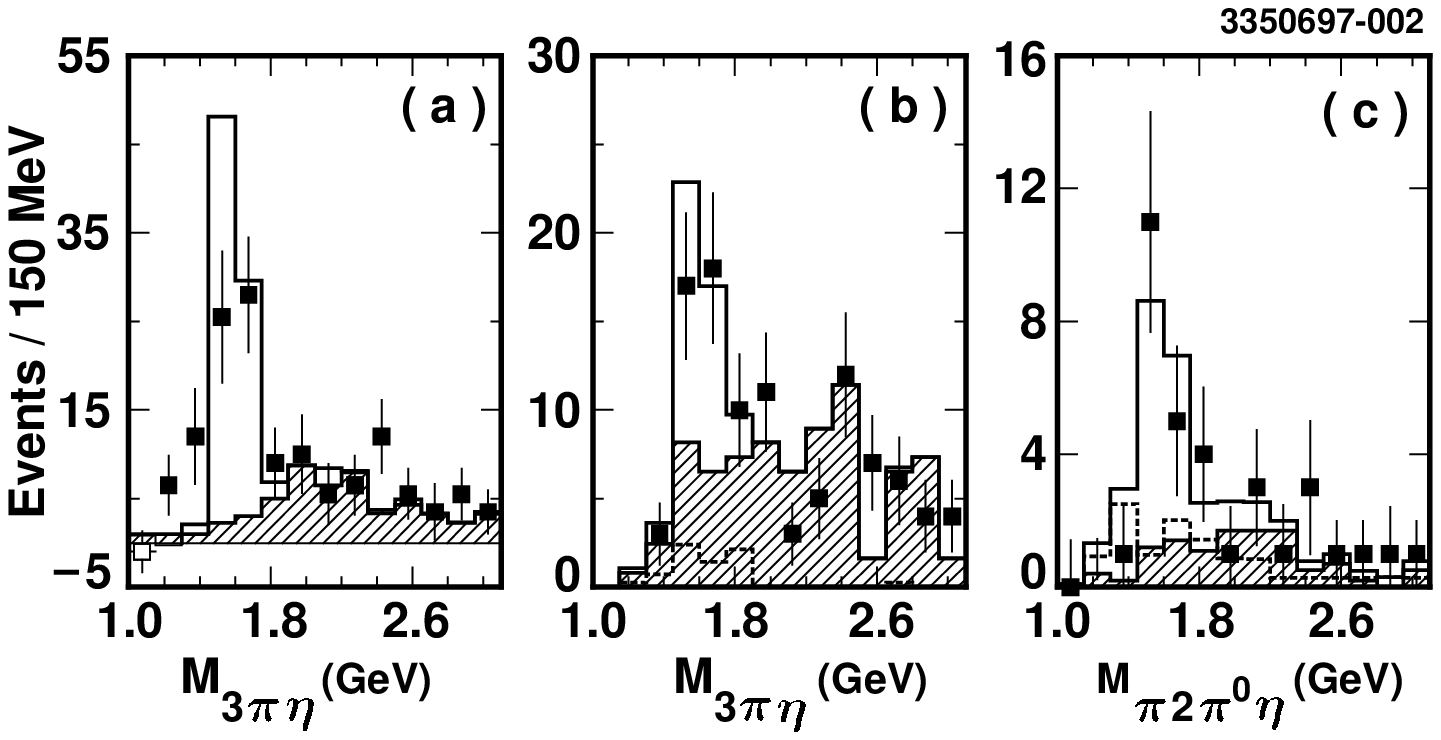}{2.5}{
Hadronic mass spectra for the (a) {\thrhgg} (after $\eta$ sideband subtraction), (b)
{\thrh3p}, (c) {\h2pgg}  samples.  The solid line is a fit to the data (squares).
The tau and $q\bar{q}$ backgrounds are indicated by the dashed line and hatched 
area, respectively. 
}

We extract the number of $\eta$'s by fitting the distributions shown in Fig. 1.
In Table \ref{tab:3heta} we present results from the three analyses.
Efficiencies shown in Tables I and II include tagging branching fractions.

\begin{table}[htbp]
\caption{\label{tab:3heta}
The number of data signal events, efficiencies, and branching fractions for the 
three data analyses.  The signal has been corrected for the background.}
\def\1#1#2#3{\multicolumn{#1}{#2}{#3}}
\begin{center}
\begin{tabular}{l c c c}
{Sample}&{N$^{\eta}_{data}$}&{$\eff$[\%]}&$\B(\times10^{-4})$\\ 
\hline 
\thrhgg             &73.4$^{+13.0}_{-12.3}$ &6.3&3.5$^{+0.7}_{-0.6}\pm$0.7 \\ 
\thrh3p             &15.2$^{+4.8}_{-4.6}$   &1.8&3.1$^{+0.9}_{-0.9}\pm$1.0 \\
\h2pgg              & 15.0$^{+5.0}_{-5.0}$   &2.5&1.4$^{+0.6}_{-0.6}\pm$0.3 \\ 
\end{tabular}
\end{center}
\end{table}               

We estimate several sources of systematic errors.  The major contributions 
are (for \thrhgg, \thrh3p, and \h2pgg\ samples):
$\piz$ and $\eta$ reconstruction efficiency (10\%, 10\%, 4\%);
model dependence (5\%, 10\%, 10\%); backgrounds (15\%, 17\%, 18\%).
The total systematic errors are 19\%, 33\%, and 20\%, respectively.

For the \thrhgg\ sample,
there is enough data for a consistency check of 
$\B(\tau\to \pi^-\pi^+\pi^-\eta\nu_{\tau})$ measurements among data samples selected 
with e, $\mu$ and hadron tags. They are consistent with each other, with a
$\chi^2$ of 5.1 for three degrees of freedom, corresponding to 16\% confidence 
level.  Combining \thrhgg\ and  \thrh3p\  results, we obtain  
$\B(\tau^{-}\to \pi^-\pi^+\pi^-\eta\nu_{\tau})=(3.4^{+0.6}_{-0.5}\pm0.6)\times10^{-4}$.

\Single{3}{t}{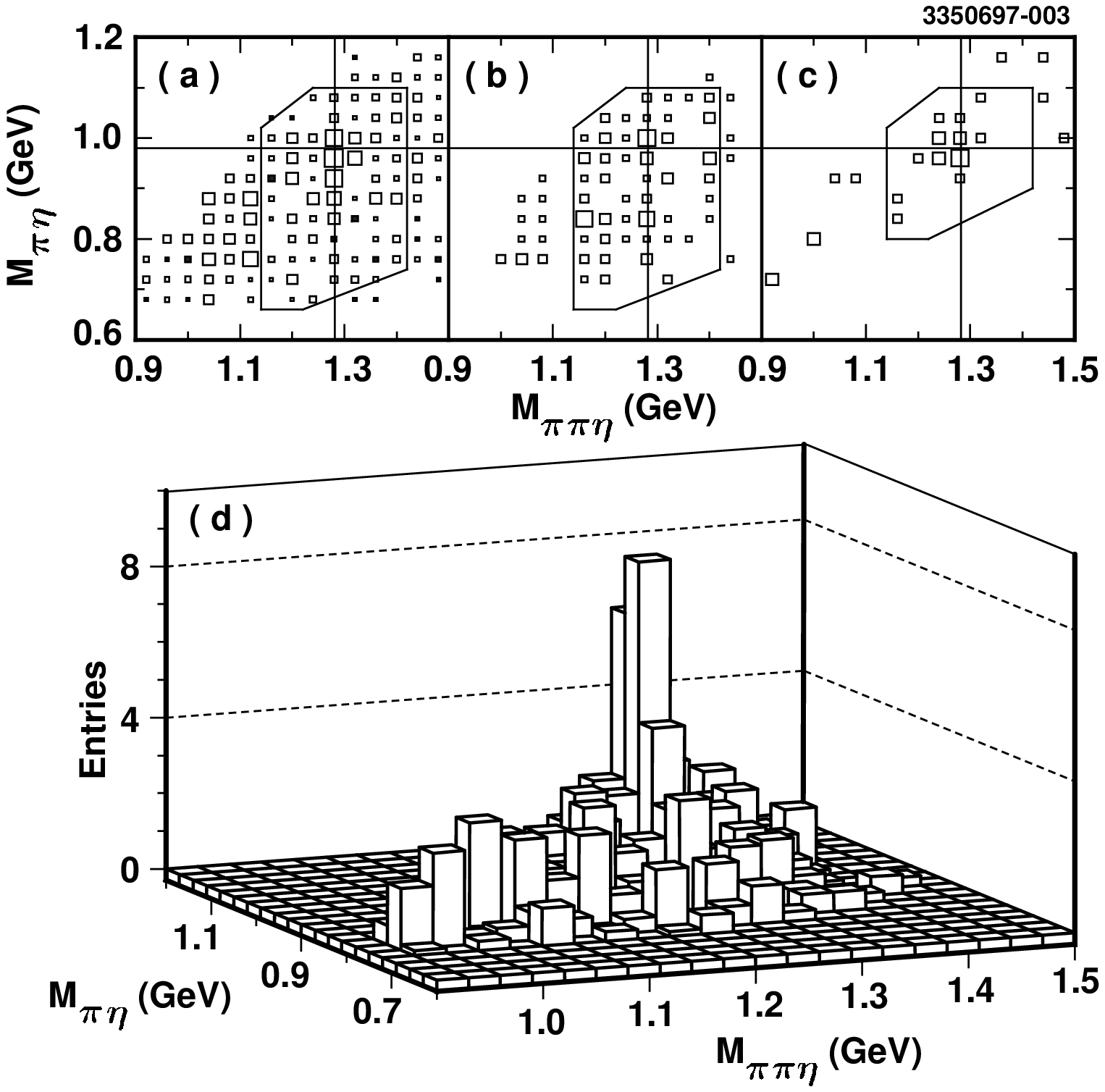}{4.2}{
M($\pi\eta$) vs M($\eta\pi\pi$) for the (a) {\thrhgg} (after $\eta$ sideband subtraction),
 (b) {\thrh3p}, (c) {\h2pgg} samples. Plot (d) is a sum of (a), (b) and (c) weighted as 
0.25, 0.25 and 1 respectively. All bin widths are 40 MeV.}

\nopagebreak

A $3\pi\eta$ final state could proceed through
a number of different resonances.
In Fig. 3  for events from  the $\eta$ signal region 
we plot the  $\pi\eta$ vs $\eta\pi\pi$ 
mass distributions, using the mass-constrained 
$\eta$ and $\piz$ momenta.  
The distributions show higher population density in the 
$f_1(1285)$/$a_0(980)$ region, indicating the presence of the decay chain: 
$\tau^-\to f_1\pi^-\nu_{\tau}$, $f_1\to a_0(980)\pi$, $a_0(980)\to\eta\pi$.
For the 3-prong modes (Figs. 3(a) and (b)) there is an ambiguity in the choice
of the same charge pions that results in four entries per event.
In the case of $\tau^{-}\to \pi^{-}2\piz\eta\nu_{\tau}$ (Fig. 3(c)), 
there is only one
$f_1\to\eta\piz\piz$ combination and two $\eta\piz$ combinations.
Since the kinematics of the $f_1\to a_0\piz$ decay are such that the higher 
mass $\eta\piz$ combination is the correct one about 90\% of the time,
we plot only the higher-mass $\eta\piz$ combination.

We expect that more than $75\%$ of all $f_1\to\eta\pi\pi$ decays 
proceed through the $a_0(980)\pi$ state\cite{PDG}.
To extract the number of the $f_1\pi$ events we perform a binned maximum 
likelihood fit.
We restrict the fit to the area shown in Fig. 3 to avoid the kinematically 
forbidden region and we weight each event by the inverse of the number of 
entries.  The fit function is the sum of a signal-MC
$a_0$ vs $f_1$  distribution, a random $\gamma\gamma$ background shape 
obtained from non-signal tau MC and a constant background. 
For the signal MC we use the full decay chain with $f_1$ and
$a_0(980)$ resonances \cite{V-A} and include all
$\eta\pi\pi$ and $\eta\pi$ mass combinations. 
The constant background accounts for a possible non-$f_1$ signal.
 To take into account the uncertainty in the random $\gamma\gamma$ 
background normalization, we perform a combined fit of the 
M$({\pi\eta})$ vs M(${\pi\pi\eta})$ and $S_{\gamma\gamma}$
 data distributions with normalizations of the random 
$\gamma\gamma$ background constrained to be the same.
From Monte Carlo studies we find that all three fits 
have confidence levels above 18\%.  We have found no sources 
of background which can contribute to the $f_1$ peak.

In Table \ref{tab:f1pi} we show the fit results obtained for the different 
data samples.  We have used $\B(f_1(1285)\to\eta\pi\pi)=0.54\pm0.15$ \cite{PDG}
and an isospin factor of 2/3 (1/3)  for $\eta\pi^+\pi^-(\eta\pi^0\pi^0)$.
We include a systematic error of 28\% to account for 
the uncertainty of the $f_1\to\eta\pi\pi$ decay rate.  All other contributions
to the total systematic error including different models of the 
$f_1\to\eta\pi\pi$ decay are found to be much smaller.  
The total systematic error is 33\% for all three channels. 

\begin{table}[htbp]
\caption{\label{tab:f1pi}
The numbers of signal events, efficiencies and branching 
fractions for the $\tau^{-}\to f_1\pi^{-}\nu_{\tau}$  decays obtained from fits.}
\def\1#1#2#3{\multicolumn{#1}{#2}{#3}}
\begin{center}
\begin{tabular}{l c c c} 
{Sample}& {N$^{f_1}_{data}$} & {$\eff$[\%]} & $\B(\times10^{-4})$ \\
\hline
\thrhgg             & 36.3$^{+9.7}_{-9.0}$ & 5.6 & 5.3$^{+1.4}_{-1.3}\pm$1.8 \\
\thrh3p             & 9.6$^{+5.6}_{-4.7}$ & 1.4 & 6.8$^{+4.0}_{-3.3}\pm$2.2 \\
\h2pgg              & 8.4$^{+3.2}_{-3.2}$ & 2.6 & 6.6$^{+2.5}_{-2.5}\pm$2.3 \\
\end{tabular}
\end{center}
\end{table}               

The weighted average for all three channels is 
$\B(\tau^{-}\to f_1\pi^{-}\nu_{\tau})=(5.8^{+1.4}_{-1.3}\pm1.8)\times10^{-4}$.
Using results from the \thrhgg\ channel we find the ratio 
$\B(\tau^{-}\to f_1\pi^{-}\nu_\tau\to\pi^-\pi^+\pi^-\eta\nu_\tau) / 
 \B(\tau^{-}\to\pi^-\pi^+\pi^-\eta\nu_\tau)=0.55\pm0.14$.
Here we take advantage of the fact that some systematic uncertainties
are canceled in the ratio. It would appear that not all of the $\pi^-\pi^+\pi^-\eta$
final state proceeds through an intermediate $f_1$.
We show in Fig. 4 the background-subtracted distribution of 
$M(\pi^-\pi^+\pi^-\eta)$ calculated for \thrhgg\ events with a
$\pi^+\pi^-\eta$ mass of at least 36 MeV from the nominal $f_1$ mass.
This distribution as well as distributions of all
other sub-mass projections are  consistent with 
$\tau^-\to a_1(1260)^-\eta\nu_{\tau}\to\pi^-\rho^0\eta\nu_{\tau}$
decay model. Since the observed excess has less than $2.0\sigma$ significance,
we set an upper limit:
$\B(\tau^-\to a_1(1260)^-\eta\nu_{\tau}\to\pi^-\rho^0\eta\nu_{\tau})<3.9\times10^{-4}$ at 90\% CL.

\Ssingle{4}{h}{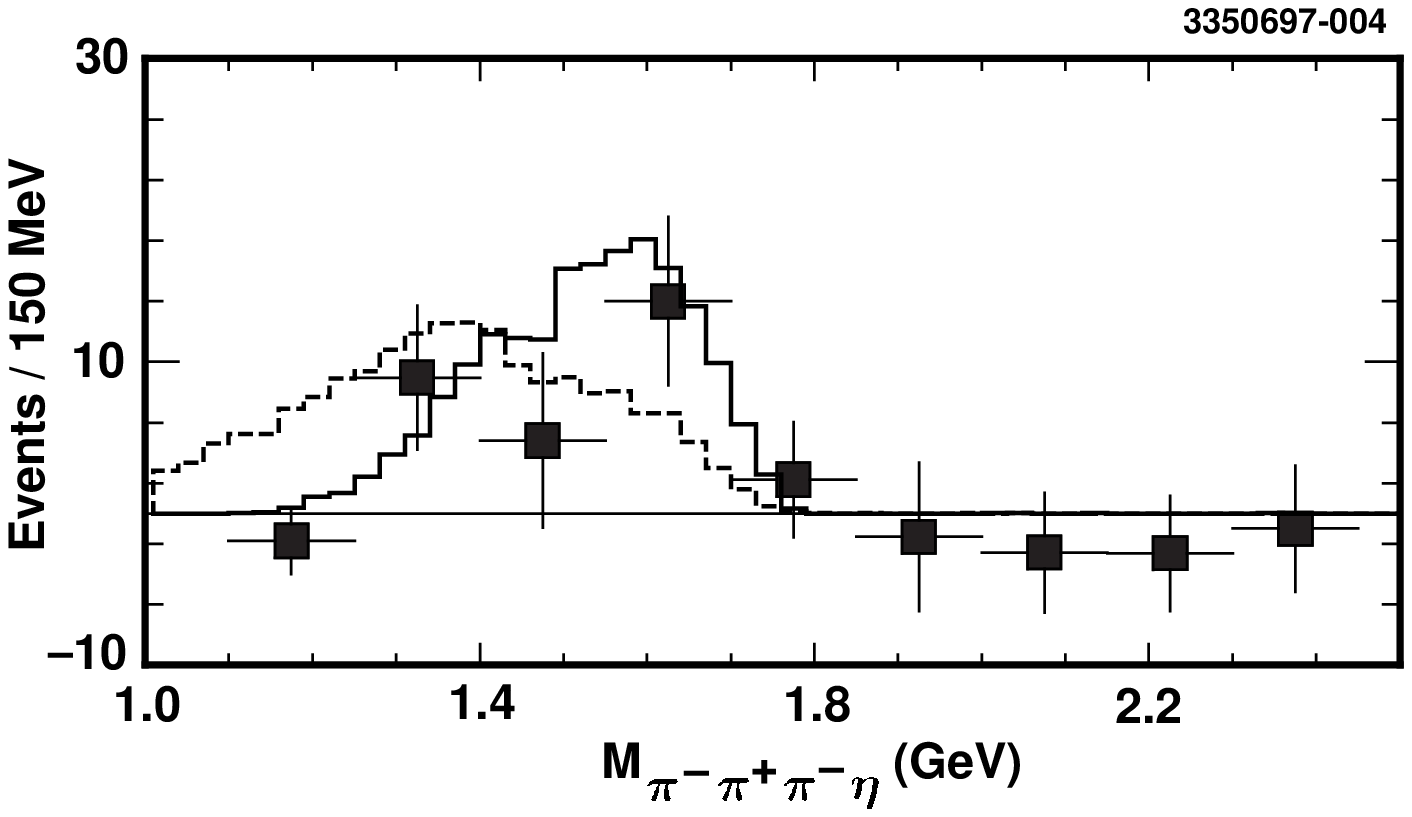}{3.2}{ Distribution of $\pi^-\pi^+\pi^-\eta$
 invariant mass plotted for \thrhgg\ events not associated with
$\tau^{-}\to f_1\pi^{-}\nu_\tau$ decay.
The Monte Carlo expectations for the decays $\tau^-\to\rho^0\pi^-\eta\nu_{\tau}$ and 
$\tau^-\to a_1(1260)^-\eta\nu_{\tau}\to\pi^-\pi^+\pi^-\eta\nu_{\tau}$
are shown as dashed and solid lines, respectively. Both Monte Carlo modes are generated
via phase space with ``V-A'' factor\cite{V-A}.}

The measured branching fraction for the decay 
$\tau^-\to \pi^-\pi^+\pi^-\eta\nu_\tau$
is more than two orders of magnitude larger than the value calculated by
Pich \cite{pich} under the assumption that this decay is dominated by an
$\eta a_1$ intermediate state.
In a recent calculation, Li \cite{li1} used similar assumptions to
obtain $\B(\tau^-\to a_1(1260)^-\eta\nu_{\tau}\to\pi^-\rho^0\eta\nu_{\tau})=2.93\times10^{-4}$.
In another paper, Li \cite{li} calculated  $\B(\tau^-\to f_1\pi^-\nu_\tau)=2.9\times10^{-4}$,
which is still somewhat smaller than the present measurement. 

The decay $\tau\to 3\pi\eta\nu_\tau$ has important
implications for the phenomenology of the multi-pion $\tau$ decays, especially 
for $\tau\to 6\pi\nu_\tau$.
Several authors [14-16] have used isospin relations \cite{pais} to calculate the
relative amounts of $\tau^-\to 3\pi^-2\pi^+\piz\nu_\tau$, $\tau^-\to
\pi^-\pi^+\pi^-3\piz\nu_\tau$ and $\tau^-\to \pi^-5\piz\nu_\tau$,
and claimed some discrepancies between the measured branching
fractions and CVC predictions obtained from the $e^+e^-\to 6\pi$ measurements.

It now appears that $\tau^-\to 3\pi^-2\pi^+\piz\nu_\tau$ and $\tau^-\to
\pi^-\pi^+\pi^-3\piz\nu_\tau$ decays have large contributions from the
$\tau\to3\pi\eta\nu_\tau$ channel.  Since this final state has opposite 
G parity to that of the direct $6\pi$ decays and proceeds through an 
axial-vector current, its contribution must be subtracted
before applying isospin relations or using CVC to compare with
$e^+e^-$ annihilation data.

\Sfingle{5}{h}{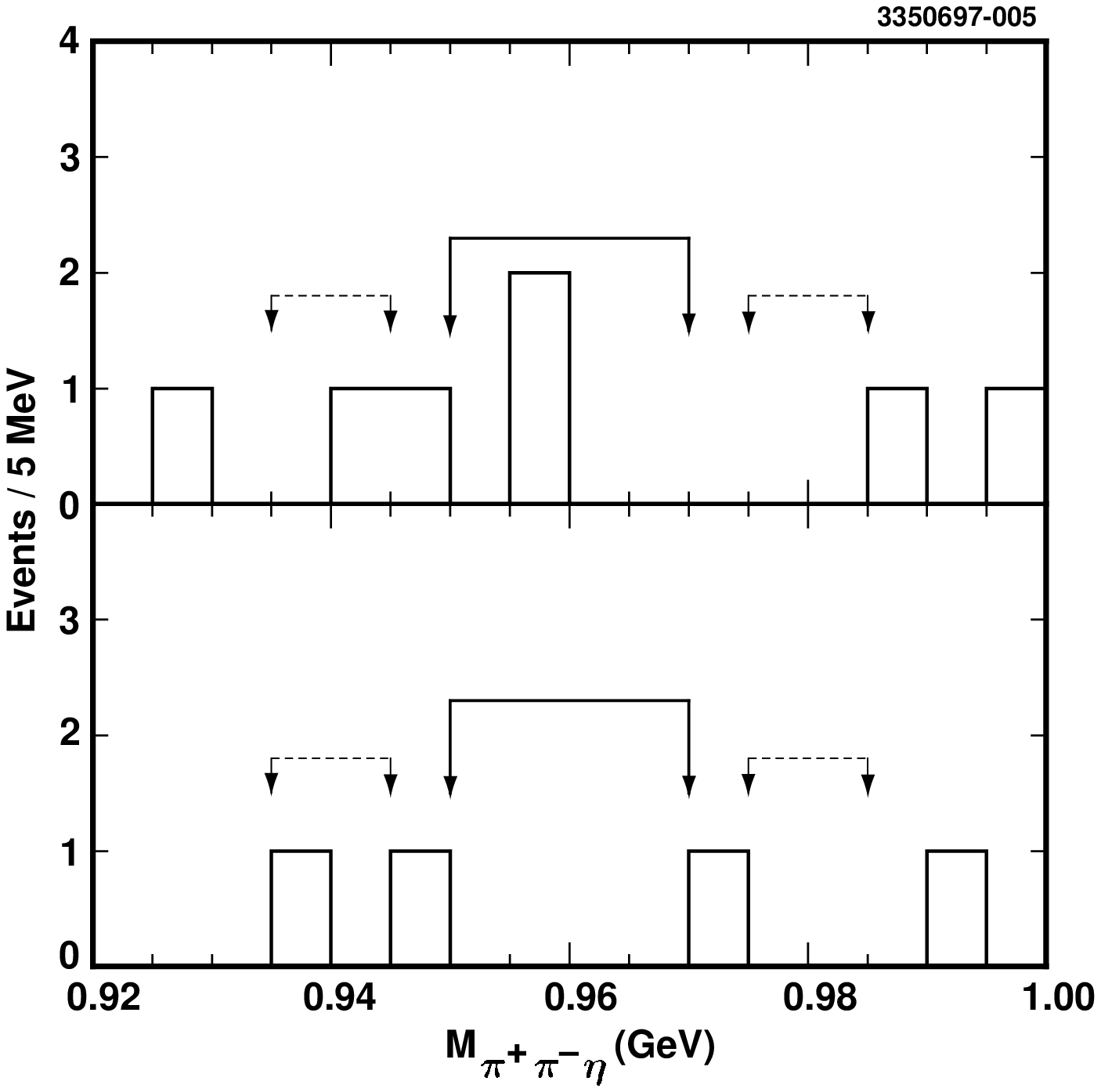}{3.5}{Distribution of the $M_{\pi^+\pi^-\eta}$
plotted for: (a) $\tau^{-}\to \pi^{-}\eta'(958)\nu_{\tau}$ and
(b) $\tau^{-}\to \pi^{-} \piz \eta'(958)\nu_{\tau}$ signal candidates.
The solid (dashed) arrows indicate the signal (sideband) regions.}

We use the selected samples of $\tau^-\to \pi^-\pi^+\pi^-\eta\nu_{\tau}$
events to set upper limits on the $\tau^-\to \pi^- \eta'(958)\nu_{\tau}$ 
and $\tau^{-}\to \pi^{-} \piz \eta'(958)\nu_{\tau}$ decays.
Each event must contain an $\eta$ candidate: $0.51$ GeV$<M_{\eta}<0.57$ GeV.
For the second decay, the remaining photons with $E_{\gamma}>30$ MeV
are used for the $\piz$ reconstruction. 
In the $M_{\pi^+\pi^-\eta}$ signal(sideband) region shown in Figs. 5(a,b) 
we find 2(1) and 0(1) events respectively. 
The event detection efficiencies are 4.4\% and 2.3\%, with relative systematic
uncertainties of 11\% and 15.6\% respectively.
Using  Poisson statistics and assuming a linear background distribution we obtain \cite{kk}:
$\B(\tau^{-}\to \pi^{-}\eta'(958)\nu_{\tau})<7.4\times10^{-5}$ and
$\B(\tau^{-}\to \pi^{-}\piz\eta'(958)\nu_\tau)<8.0\times10^{-5}$.

We gratefully acknowledge the effort of the CESR staff in providing us with
excellent luminosity and running conditions.
This work was supported by 
the National Science Foundation,
the U.S. Department of Energy,
the Heisenberg Foundation,  
the Alexander von Humboldt Stiftung,
Research Corporation,
the Natural Sciences and Engineering Research Council of Canada,
and the A.P. Sloan Foundation.


\begin{thebibliography}{99}
\bibitem{charge}{Throughout this paper charge conjugation is implied.}
\bibitem{Mmiss}{$M_{miss}=\sqrt{(E_{cm}-E_{vis})^2-P_{vis}^2}$,
where $E_{vis}$ and $P_{vis}$ are the total visible energy 
and momentum of the event.}
\bibitem{V-A}{
Monte Carlo  events are  generated according to phase space  with a ``V-A''
factor, $(M_\tau^2-q^2)^2(M_\tau^2+2q^2)$,
and a P-wave barrier factor $8p_\pi^3/q^3$,
where $p_{\pi}$ is the momentum of the pion in the $f_{1}\pi$ center of mass
and $q$  is the total effective mass of the hadrons.
We assume the decay spectral function to be dominated by the 
form factor of the $a_{1}$(1260) resonance.}
\end{thebibliography}
\end{document}